\documentclass[prl,twocolumn,superscriptaddress,showpacs,aps]{revtex4} 
\usepackage{graphicx} 
\begin{document} 
\newcommand{\sigm}{$\Sigma^{-}$} 
\newcommand{\sigs}{$\Sigma^{*-}$} 
\title{Cross Sections and Beam Asymmetries for $K^+\Sigma^{*-}$ photoproduction 
from the deuteron at $E_\gamma=$1.5-2.4 GeV} 
 
\newcommand*{\OHIOU}{Department of Physics and Astronomy, Ohio University, 
	Athens, Ohio 45701, USA} 
\newcommand*{\RCNP}{Research Center for Nuclear Physics, Osaka University,
	Ibaraki, Osaka 567-0047, Japan} 
\newcommand*{\PUSAN}{Department of Physics, Pusan National University,
	Busan 609-735, Korea} 
\newcommand*{\KONAN}{Department of Physics, Konan University,
	Kobe, Hyogo 658-8501, Japan} 
\newcommand*{\KPSI}{Kansai Photon Science Institute, Japan Atomic Energy Agency,
	Kizu, Kyoto 619-0215, Japan} 
\newcommand*{\SINICA}{Institute of Physics, Academia Sinica,
	Taipei, Taiwan 11529, Republic of China} 
\newcommand*{\SP}{Japan Synchrotron Radiation Research Institute, 
	Mikazuki, Hyogo 679-5198, Japan} 
\newcommand*{\NAGOYA}{Department of Physics and Astronomy, Nagoya University, 
	Nagoya, Aichi 464-8602, Japan} 
\newcommand*{\SEOUL}{School of Physics, Seoul National University, 
	Seoul 151-747, Korea} 
\newcommand*{\KYOTO}{Department of Physics, Kyoto University, 
	Kyoto 606-8502, Japan} 
\newcommand*{\TOHOKU}{Laboratory of Nuclear Science, Tohoku University, 
	Sendai, Miyagi 982-0826, Japan} 
\newcommand*{\YAMA}{Department of Physics, Yamagata University, 
	Yamagata 990-8560, Japan} 
\newcommand*{\CHIBA}{Department of Physics, Chiba University,
	Chiba 263-8522, Japan} 
\newcommand*{\WAKA}{Wakayama Medical College, Wakayama 641-8509, Japan} 
\newcommand*{\MIYA}{Department of Applied Physics, Miyazaki University,
	Miyazaki 889-2192, Japan} 
\newcommand*{\TOKYO}{Department of Physics, Tokyo Institute of Technology, 
	Tokyo 152-8551, Japan} 
\newcommand*{\OSAKA}{Department of Physics, Osaka University,
	Toyonaka, Osaka 560-0043, Japan} 
\newcommand*{\TAMU}{Cyclotron Institute and Physics Department, 
	Texas A\&M University, College Station, Texas, 77843, USA}
\newcommand*{\SASK}{Department of Physics, University of Saskatchewan,
	Saskatoon, Saskatchewan, Canada} 
\newcommand*{\MINN}{School of Physics and Astronomy, University of Minnesota,
	Minneapolis, Minnesota 55455, USA} 
\newcommand*{\AKITA}{Akita Research Institute of Brain and Blood Vessels, 
	Akita 010-0874, Japan} 
\newcommand*{\IPR}{Institute for Protein Research, Osaka University, 
	Suita, Osaka 565-0871, Japan} 
\newcommand*{\MSU}{NSCL, Department of Physics and Astronomy, 
	Michigan State University, East Lansing, MI 48824, USA} 
\author {K.~Hicks}  
\affiliation{\OHIOU} 
\author {D.~Keller}  
\affiliation{\OHIOU} 
\author {H.~Kohri}  
\affiliation{\RCNP} 
\author {D.S.~Ahn}  
\affiliation{\RCNP}
\affiliation{\PUSAN} 
\author {J.K.~Ahn}  
\affiliation{\PUSAN} 
\author {H.~Akimune}  
\affiliation{\KONAN} 
\author {Y.~Asano}  
\affiliation{\KPSI} 
\author {W.C.~Chang}  
\affiliation{\SINICA} 
\author {S.~Dat\'e}  
\affiliation{\SP} 
\author {H.~Ejiri}  
\affiliation{\RCNP}
\affiliation{\SP} 
\author {S.~Fukui}  
\affiliation{\NAGOYA} 
\author {H.~Fujimura}  
\affiliation{\SEOUL}
\affiliation{\KYOTO} 
\author {M.~Fujiwara}  
\affiliation{\RCNP}
\affiliation{\KPSI} 
\author {S.~Hasegawa}  
\affiliation{\RCNP} 
\author {T.~Hotta}  
\affiliation{\RCNP} 
\author {K.~Imai}  
\affiliation{\KYOTO} 
\author {T.~Ishikawa}  
\affiliation{\TOHOKU} 
\author {T.~Iwata}  
\affiliation{\YAMA} 
\author {Y.~Kato}  
\affiliation{\RCNP} 
\author {H.~Kawai}  
\affiliation{\CHIBA} 
\author {Z.Y.~Kim}  
\affiliation{\SEOUL} 
\author {K.~Kino}  
\affiliation{\RCNP} 
\author {N.~Kumagai}  
\affiliation{\SP} 
\author {S.~Makino}  
\affiliation{\WAKA} 
\author {T.~Matsuda}  
\affiliation{\MIYA} 
\author {T.~Matsumura}  
\affiliation{\RCNP}
\affiliation{\SP} 
\author {N.~Matsuoka}  
\affiliation{\RCNP} 
\author {T.~Mibe}  
\affiliation{\RCNP}
\affiliation{\OHIOU} 
\author {M.~Miyabe}  
\affiliation{\KYOTO} 
\author {Y.~Miyachi}  
\affiliation{\TOKYO} 
\author {M.~Morita}  
\affiliation{\RCNP} 
\author {N.~Muramatsu}  
\affiliation{\SP}
\affiliation{\RCNP} 
\author {T.~Nakano}  
\affiliation{\RCNP} 
\author {M.~Niiyama}  
\affiliation{\KYOTO} 
\author {M.~Nomachi}  
\affiliation{\OSAKA} 
\author {Y.~Oh}  
\affiliation{\TAMU} 
\author {Y.~Ohashi}  
\affiliation{\SP} 
\author {H.~Ohkuma}  
\affiliation{\SP} 
\author {T.~Ooba}  
\affiliation{\CHIBA} 
\author {J.~Parker}  
\affiliation{\KYOTO} 
\author {C.~Rangacharyulu}  
\affiliation{\SASK} 
\author {A.~Sakaguchi}  
\affiliation{\OSAKA} 
\author {T.~Sasaki}  
\affiliation{\KYOTO} 
\author {P.M.~Shagin}  
\affiliation{\MINN} 
\author {Y.~Shiino}  
\affiliation{\CHIBA} 
\author {A.~Shimizu}  
\affiliation{\RCNP} 
\author {H.~Shimizu}  
\affiliation{\TOHOKU} 
\author {Y.~Sugaya}  
\affiliation{\OSAKA} 
\author {M.~Sumihama}  
\affiliation{\OSAKA}
\affiliation{\SP} 
\author {Y.~Toi}  
\affiliation{\MIYA} 
\author {H.~Toyokawa}  
\affiliation{\SP} 
\author {A.~Wakai}  
\affiliation{\AKITA} 
\author {C.W.~Wang}  
\affiliation{\SINICA} 
\author {S.C.~Wang}  
\affiliation{\SINICA} 
\author {K.~Yonehara}  
\affiliation{\KONAN} 
\author {T.~Yorita}  
\affiliation{\RCNP}
\affiliation{\SP} 
\author {M.~Yoshimura}  
\affiliation{\IPR} 
\author {M.~Yosoi}  
\affiliation{\KYOTO}
\affiliation{\RCNP} 
\author {R.G.T.~Zegers}  
\affiliation{\MSU} 
\collaboration{The LEPS Collaboration} 
\noaffiliation 
\date{\today\\} 
 
\begin{abstract} 
The $\Sigma(1385)$ resonance, or $\Sigma^*$, is well-known as part of 
the standard baryon decuplet with spin $J=3/2$. 
Measurements of the reaction $\gamma p \to K^+ \Sigma^{*0}$ are difficult 
to extract due to overlap with the nearby $\Lambda(1405)$ resonance.  
However, the reaction $\gamma n \to K^+ \Sigma^{*-}$ has no overlap 
with the $\Lambda(1405)$ due to its charge.
Here we report the first measurement of cross sections and 
beam asymmetries for photoproduction of the \sigs~ from a deuteron target, 
where the $K^+$ and $\pi^-$ are detected in the LEPS spectrometer.  
The cross sections at forward angles range from 0.4 to 1.2 $\mu$b, 
with a broad maximum near $E_\gamma \simeq 1.8$ GeV. 
The beam asymmetries are negative, in contrast to postive values 
for the $\gamma n \to K^+\Sigma^-$ reaction.
\end{abstract} 
 
 
\maketitle 
 
The photoproduction of the spin $J=3/2$ baryon resonance with strangeness 
$-1$ at a mass of 1385 MeV, known as the $\Sigma^*$, is virtually unknown 
except for a few early bubble-chamber experiments \cite{bubble}, 
preliminary data from CLAS \cite{guo}, and data 
from LEPS with large photon energy bins \cite{niiyama}
(all from a proton target).  
The $\Sigma^*$ is the lowest mass strange 
baryon in the decuplet of $J=3/2$ baryons, so measurements of 
its production cross section and spin observables are of intrinsic interest to 
compare with theoretical calculations based on SU(3) symmetry 
and the well known $\Delta$ resonance photoproduction amplitudes
\cite{close}. 

A recent paper by Oh, Ko and Nakayama \cite{oh} 
states ``photoproduction of $K\Sigma^*$ provides a useful tool for 
testing baryon models".  In particular, the $K\Sigma^*$ final state, 
produced at higher $W$ than for ground-state $K\Lambda$ or $K\Sigma$ 
production, may be a good way to search for the so-called ``missing" 
resonances (those not listed by the Particle Data Group \cite{pdg})
as predicted by the constituent quark model \cite{capstick}.
For example, D\"oring {\it et al.}~\cite{doring} suggest that the 
$\Delta(1700)$ resonance may have a large coupling to the $K\Sigma^*$ 
channel.  There are many measurements of $\Delta$ production, 
due to its strong production cross section, but this does not 
give information on the $K\Sigma^*$ coupling.

Comparison between data for $K\Sigma^*$ production and 
reaction models is one way to search for new resonances.
For example, in Ref. \cite{oh}, preliminary data \cite{guo} for the 
reaction $\gamma p \to K^+ \Sigma^{*0}$ near threshold are better 
fitted, within this effective Lagrangian model, with the addition of 
several high-mass resonances (see \cite{oh} for details).
However, a larger body of data including spin asymmetry observables 
is needed before any conclusions can be reached on the need for 
additional baryon resonances.

Precise data for photoproduction of the $\Sigma^*$ resonance 
will also help to constrain the SU(3) relations in production 
mechanisms of decuplet baryons.  
In the model of Ref. \cite{oh}, many coupling constants are 
not constrained by data, so standard SU(3) flavor symmetry relations 
\cite{pdg} were used for these coupling constants ({\it e.g.} those
relating the $\pi N\Delta$ and $K N\Sigma^*$ couplings).
The magnitude of SU(3) flavor symmetry breaking 
in decuplet baryon production is most easily tested 
by comparing theoretical models with data
for both $\Delta$ and $\Sigma^*$ photoproduction.
As mentioned above, little data is available for production 
of decuplet baryons beyond the $\Delta$, leaving SU(3) 
flavor symmetry largely untested among these baryons.

Here we report on the first photoproduction measurement of the 
$K^+ \Sigma^{*-}$ final state, done at the SPring-8 facility with 
the LEPS detector \cite{sumihama} and a deuterium target. 
Both the $K^+$ and a coincident $\pi^-$ 
(from the strong decay $\Sigma^{*-} \to \pi^- \Lambda$) were detected.  
Photons in the energy range of 1.5 to 2.4 GeV were produced using 
Compton backscattering of polarized laser light from 8 GeV electrons 
in the SPring-8 storage ring.  Details of the beam and detectors 
have been described previously \cite{sumihama}.  Briefly, the 
photon beam is incident on a 16 cm liquid deuterium target, 
followed by a scintillator called the start counter (SC) and  
a Cerenkov detector to veto $e^+e^-$ pairs.
Charged particles go through tracking 
detectors, into a dipole magnet, through wire chambers and on to a 
time-of-flight (TOF) scintillator wall placed 4 m from the target. 
The photon energy is calculated from the struck electron, which 
goes into a tagging spectrometer inside the SPring-8 storage ring.
The trigger is formed from a tagged electron, the SC
with no Cerenkov veto, and a hit in the TOF wall.  The TOF resolution 
was about 150 ps, and the typical momentum resolution was 6 MeV/c 
at a momentum of 1 GeV/c.

\begin{figure}[t]
\includegraphics[scale=0.4]{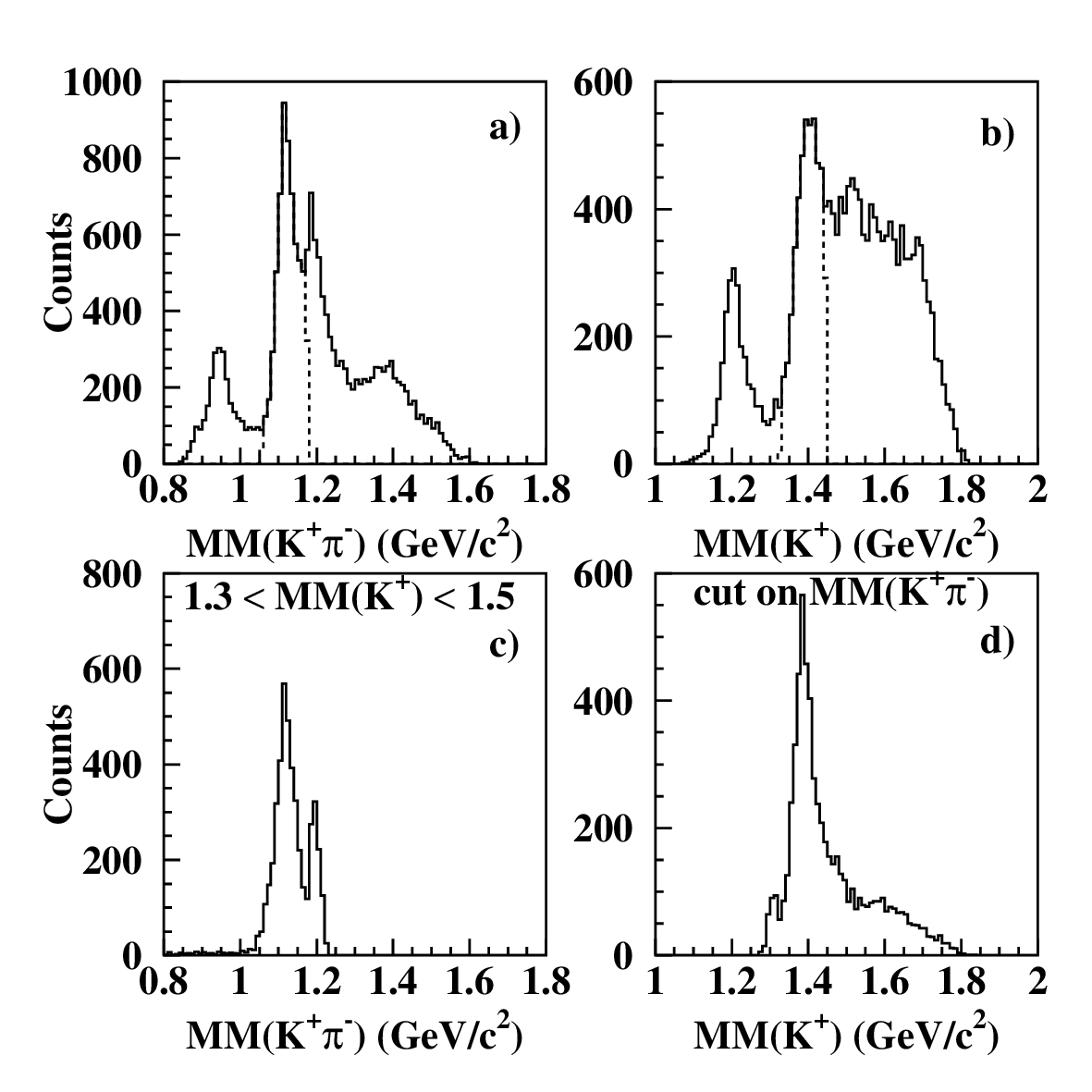} 
\caption{\label{fig:mass} 
Mass spectra calculated using the missing mass technique for:
(a) detected $K^+\pi^-$; (b) detected $K^+$ only; 
(c) same as (a) but cut on the $\Sigma(1385)$ peak shown by the 
dashed lines in (b);
(d) same as (b) but cut on the $\Lambda$ peak shown by the dashed 
lines in (a);
} 
\end{figure} 

The $\Sigma^*$ was identified using the missing mass technique.
The $K^+$ missing mass, $MM(K^+)$, is selected between 1.325 and 
1.445 GeV to isolate the \sigs~final state. For these events, 
the missing mass of the $K^+\pi^-$ system, $MM(K^+\pi^-)$, was 
fitted (using a Gaussian shape) at the $\Lambda$ mass.  The missing 
mass spectra, integrated over all angles and photon energies, 
are shown in Fig. \ref{fig:mass}, where the top 
two plots are for all events and the bottom plots are cut on the 
\sigs~and $\Lambda$ final states, respectively.  Peaks for the 
$\Sigma$ and $\Sigma^*$ final states are clearly seen in the 
uncut $MM(K^+)$ spectra (Fig. 1b). The \sigm~decays with nearly 
100\% branching ratio to $n\pi^-$, whereas the \sigs~decays 
about 87\% to $\Lambda\pi^-$ and 11.7\% to $\Sigma\pi$. 

Peaks at the neutron, $\Lambda$ and $\Sigma$ masses are clearly 
seen in the uncut $MM(K^+\pi^-)$ spectra (Fig. 1a). When a cut 
on the \sigs~peak is applied (dashed lines in Fig. 1b), the $MM(K^+\pi^-)$ 
spectra show peaks at both $\Lambda$ and $\Sigma$ masses (Fig. 1c).  
The $\Sigma$ peak near 1.2 GeV is visible in part due to
$\gamma p \to K^+\Sigma^{*0}$ production followed by the decay
$\Sigma^{*0} \to \Sigma^+\pi^-$. 

Similarly, when a cut on the $\Lambda$ 
peak is applied (dashed lines in Fig. 1a), the $\Sigma(1385)$ dominates 
the $MM(K^+)$ spectra (Fig. 1d).  The background at higher mass 
(1.5-1.8 GeV) is due to a combination of high-mass $\Sigma^*$ 
states (having greater widths) and some $\Lambda^*$ resonances 
produced from the proton, then decaying to the $\Sigma^+\pi^-$. 
Simulations show that the detector acceptance of the $\Lambda^*$ 
(after smearing for Fermi motion) is much 
smaller in the region of interest than that of the \sigs.

In the final analysis, a data cut was also applied 
to remove the small $K^{*0}$ peak in the invariant mass of the 
$K^+\pi^-$ system (not shown).  The $K^*$ production kinematics 
did not overlap with the \sigs~kinematics due to the forward 
angle detection of both decay particles in the LEPS spectrometer.

Background from $\pi^+$ particles that were misidentified as a $K^+$ 
was eliminated by calculating the missing mass $MM(\pi^+\pi^-)$ 
and then removing those events with $MM(\pi^+\pi^-)$ at the nucleon mass.

The complete set of event selections is: $MM(K^+\pi^-)$ around 
the $\Lambda$ peak; $M(K^+\pi^-)$ not at the $K^{*0}$ peak; 
and $MM(\pi^+\pi^-)$ above the nucleon mass.

Other background from the proton contribution in the deuterium 
target ({\it e.g.} due to higher-mass resonance decays) in the 
region of interest was measured with a liquid hydrogen target. 
This background (about 20\% of the signal) was directly subtracted.

The mass spectra are divided up into nine $E_\gamma$ bins 
from 1.5 to 2.4 GeV, and four bins in the kaon center-of-mass 
angle, $\theta_{cm}$. Counts for both \sigm~and \sigs~final 
states were extracted by peak fits.  The \sigm~peak had 
extremely little background (after a cut on the neutron peak in 
the $MM(K^+\pi^-)$ spectrum).  After subtracting background 
from the hydrogen target, the \sigs~decay to $\Lambda\pi^-$ 
was separated from the background leading to $\Sigma^0\pi^-$ 
(as shown in Fig. 1c) by fitting to a Gaussian with a fixed mass 
and a fixed width (extracted from Monte Carlo simulations).
The systematic uncertainty in peak fitting is estimated at 
7-8\% in each bin based on various peakshapes and polynomials 
used for the background shape. The fitted $\chi^2$ was normally 
distributed and centered near unity.

Monte Carlo simulations were carried out using the 
GEANT based \cite{geant} {\it g3leps} software \cite{sumihama}. 
Events for final states of $K^+\Sigma^{*-}$, $K^+\Sigma^{*0}$, 
$K^+\Sigma^-$ and $K^+\Lambda(1520)$ from a deuteron target were generated.
A realistic beam distribution on target was used, along with 
known detector resolutions. Fermi broadening of the initial state 
nucleon in deuterium was done by using the Paris potential.  
The photon beam energy was calculated from Compton backscattering 
kinematics, smeared by the energy width of the stored electron beam. 
Over 10 million events were generated into the detector acceptance, 
giving hundreds of events for each final state in each of the 
energy-angle bins. The event generators assumed a flat angular 
distribution in the center-of-mass frame, which is justified 
in the end because of the nearly flat angular dependence of 
the data. The counts from peak fits in each bin were divided 
by the detector acceptance from the Monte Carlo simulations, 
typically between 2\% to 9\% (depending on the bin), 
giving the unnormalized yield.

The normalization of the data used several factors (see Ref. 
\cite{sumihama} for details of the procedures).
The 16 cm LD$_2$ target had $6.77\times 10^{23}$ 
deuterons/cm$^2$. The tagger detected about $4.0\times 10^{12}$ 
photons, distributed in energy as expected from Compton scattering 
from the three UV-laser lines.  Because of material between 
the scattering point and the target \cite{sumihama}, 
only 52.6\% $\pm$ 3\% of these photons were transmitted to the target.  
A self-consistent cross check was done by comparing to previously 
published differential cross sections for the $\gamma n \to K^+ \Sigma^-$ 
reaction \cite{kohri}, which agreed to within 5\%, and is 
taken as the systematic uncertainty in the acceptance. 

\begin{figure}[t]
\includegraphics[scale=0.45]{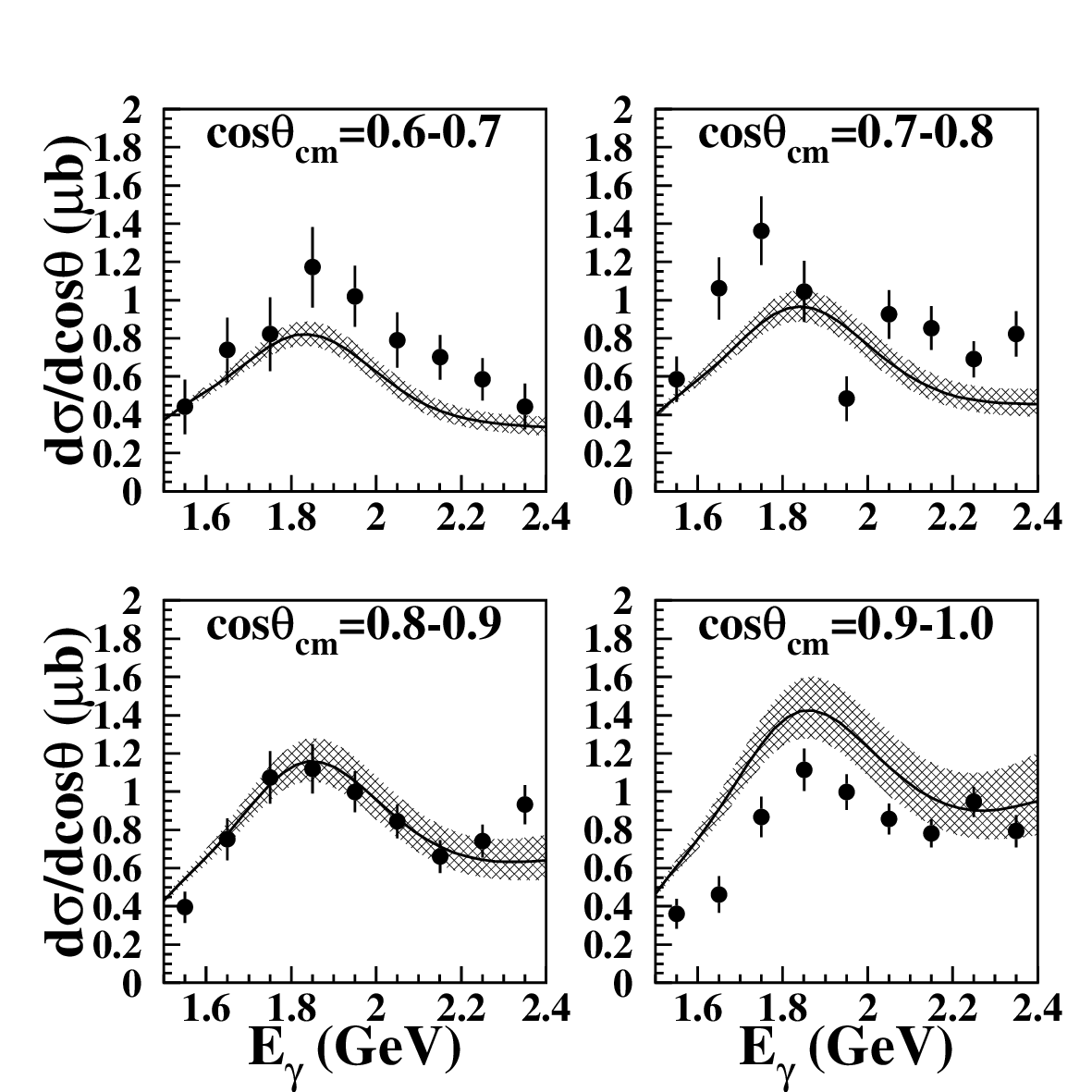} 
\caption{\label{fig:xsec} 
Cross sections for the reaction $\gamma n \to K^+\Sigma^{*-}$
as a function of photon energy.
Each plot is for the angle bin in $\cos\theta_{cm}$ as shown.
The curve is from the model of Ref. \protect\cite{oh} 
and the shaded region extends over the given angular range. 
} 
\end{figure}

Using this normalization, the yield for \sigs~production from 
the neutron (including the 87\% branching ratio for decay 
to $\Lambda\pi^-$) was converted to the cross sections shown 
in Fig. \ref{fig:xsec}.  The cross sections are, in general, 
nearly constant for a given photon energy as a function of 
$\theta_{cm}$.  As a function of photon energy, the cross sections 
initially rise, peaking at about 1.85 GeV, then fall slightly. 
The cross sections are, on average, slightly smaller than those for 
\sigm~production at the same angles \cite{kohri}. 

Systematic uncertainties are dominated by 
the peak fits and the background subtraction (from the proton), 
giving an overall systematic uncertainty of 12\%.

Theoretical calculations by Oh, Ko and Nakayama \cite{oh} 
are also shown in Fig. \ref{fig:xsec} by the solid lines
that are obtained by using the phototransition amplitude 
of resonances from the neutron as predicted in Ref. \cite{capstick}.  
The data are in reasonable agreement with the model, 
but have an angular dependence 
that is less steep than for the calculations.
In the model of Ref. \cite{oh}, the cross sections at forward 
angles are dominated by $K^+$ exchange in the $t$-channel, 
and the contribution from $K^*$ exchange is negligible. 
Since $K^*$ exchange makes a strong contribution to the 
$\gamma p \to K^+ \Lambda$ reaction, the inferred lack 
of $K^*$ exchange here will allow greater sensitivity to the 
$s$-channel diagrams in the model of Ref. \cite{oh}.
Further theoretical investigations of $N^*$ resonances that 
couple to this channel are desired.

\begin{figure}[t] 
\includegraphics[scale=0.45]{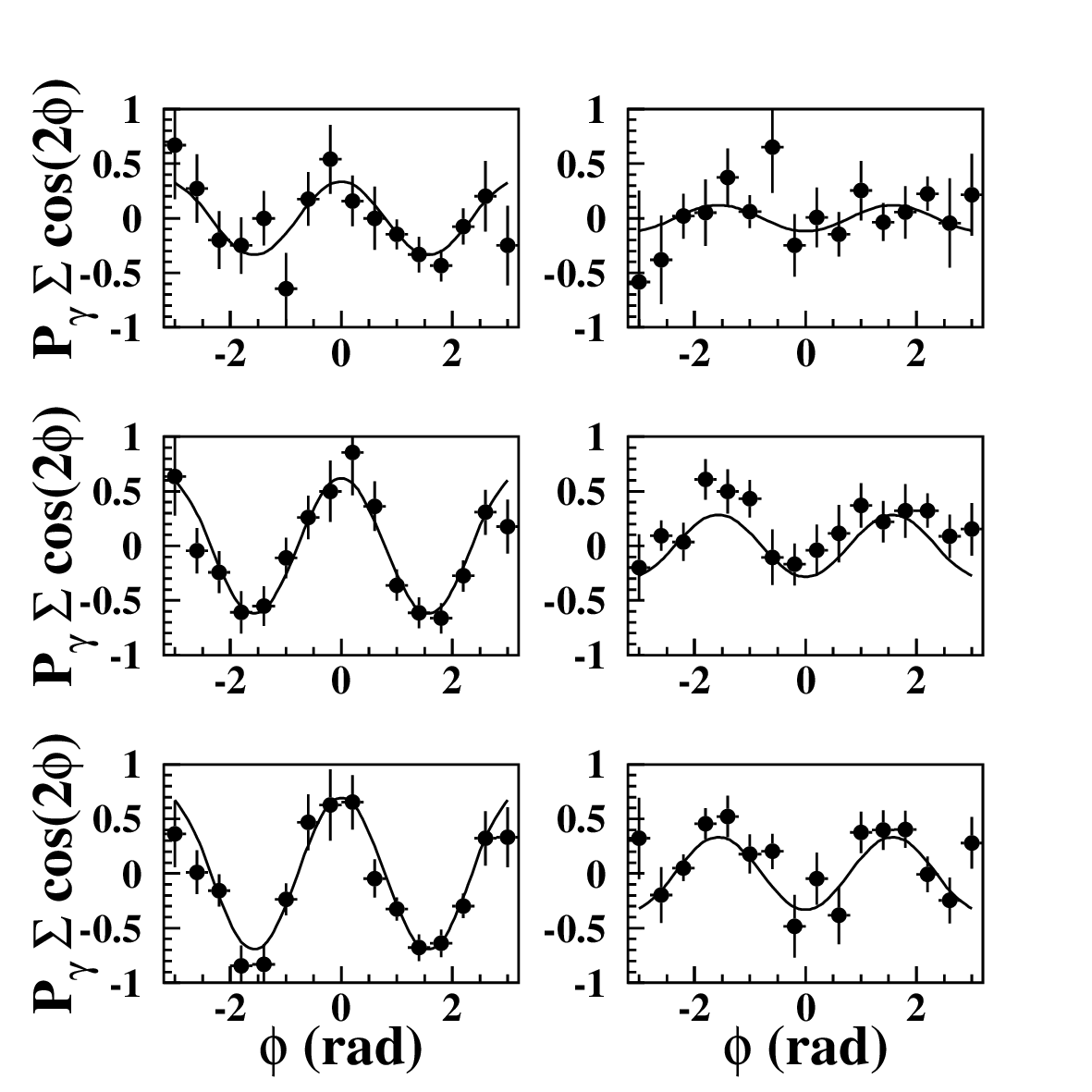} 
\caption{\label{fig:phi} 
Azimuthal angle $\phi$-dependence of the asymmetry given by 
Eq. (1) for $K^+ \Sigma^-$ (left) 
and $K^+ \Sigma^{*-}$ (right) photoproduction from the deuteron.
Three energy bins are shown:
1.5-1.8 GeV (top), 1.8-2.1 GeV (middle) and 2.1-2.4 GeV (bottom).
The curves are fit to the function $A \cos 2\phi$.
} 
\end{figure}

Another independent constraint on the theoretical models is 
given by the photon beam asymmetry, $\Sigma$.  
For the present results, 
the beam is linearly polarized, ranging from about 60\% at 
the lowest energy up to $\sim$96\% at the Compton edge.  The beam 
asymmetries can be extracted from the dependence of the cross 
section on the azimuthal angle, $\phi$, as 
shown in Fig. \ref{fig:phi}.  The beam asymmetry is defined as
\begin{equation}
 P_\gamma \Sigma \cos 2\phi = \frac{xN_v-N_h}{xN_v+N_h} 
\end{equation}
where $P_\gamma$ is the magnitude of the beam polarization, 
$N_v$ and $N_h$ are the acceptance-corrected counts for 
the beam polarized vertically and horizontally, respectively,
and $x$ is a normalization factor to account for the ratio 
of luminosity between the two beam polarizations 
($x=0.975 \pm 0.02$ for this analysis). 
The angle $\phi$ is measured between the horizontal plane 
and the reaction plane.  Contamination from background 
was removed using the same procedure as described in 
Ref. \cite{sumihama}.  The systematic uncertainty associated with 
removal of the background beam asymmetry is only a few percent.

Results for both \sigm~and \sigs~final states are shown, 
for 3 bins in energy (averaged over all angle bins).  
Superimposed are curves fit to the expected $\cos 2\phi$ 
shape.  The fit reproduces the data well, 
where only the amplitude is allowed to vary. 
The beam asymmetries are clearly much larger for \sigm~than 
for \sigs~production, and even changes sign.  
Since asymmetries may be sensitive to the structure of the 
production amplitudes, the observed small asymmetries 
constrains the main production mechanism.  It is possible 
that there are multiple contributions to the reaction 
mechanism ({\it e.g.}, several $s$-channel resonances).

\begin{figure}[t] 
\includegraphics[scale=0.4]{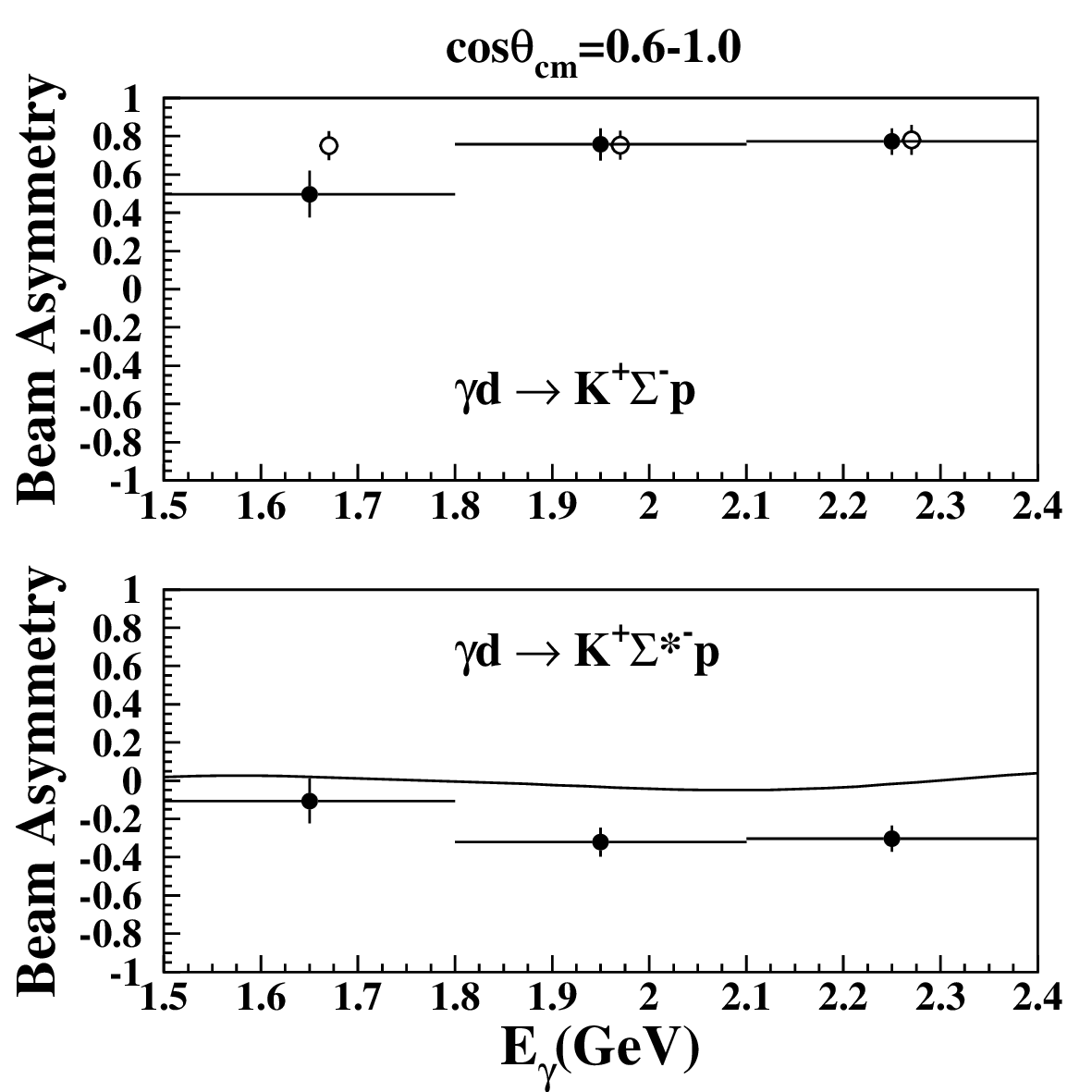} 
\caption{\label{fig:asym} 
Linear beam asymmetry, $\Sigma$, for the reactions 
$K^+ \Sigma^-$ (top) and $K^+ \Sigma^{*-}$ (bottom) 
photoproduction from the deuteron as a function of 
photon energy.  The open points in the top plot 
are weighted averages from Ref. \protect\cite{kohri}, 
which are offset from the bin center for visibility.
The curve is from the model of Ref. \protect\cite{oh}.
} 
\end{figure}

The amplitude for the fits shown in Fig. \ref{fig:phi}, 
when divided by $P_\gamma$ for the given photon energy bin, 
gives the linear beam asymmetry, $\Sigma$, as shown 
in Fig. \ref{fig:asym}.  
The large energy bins are necessary in order to get sufficient 
statistics for the fits shown in Fig. \ref{fig:phi}.  
Fits with smaller photon energy bins were also carried out, 
and no significant energy dependence was seen. 
For the \sigm~final state, the beam
asymmetries are in good agreement with the angle-averaged values 
measured in Ref. \cite{kohri} shown by the open points. 
The beam asymmetries for the \sigs~final state clearly have 
opposite sign from that of the \sigm.  
It also shows a weak energy dependence,
although more statistics are desirable to reach a firm conclusion.  
The systematic uncertainty in the beam asymmetries is about 5\%, 
primarily due to uncertainty measured for the laser polarization.

In contrast to the large values of the beam asymmetry 
for \sigm~photoproduction predicted by the KaonMAID model 
\cite{mart}, the \sigs~production model of Ref. \cite{oh} 
predicts very small beam asymmetries. It was suggested 
in Ref. \cite{oh} that the beam asymmetry is sensitive to 
the presence of $s$-channel resonance contributions for 
$\Sigma^{*0}$ photoproduction, switching from slightly 
positive for no resonances to slightly negative when all 
resonance contributions are included (note the change in 
sign in the definition of beam asymmetry for Ref. \cite{oh}).  
Calculations in this theoretical model are shown for the 
present reaction (for $\Sigma^{*-}$) by the solid line 
in Fig. \ref{fig:asym}.  These calculations have been 
averaged over the same angular range as for the current 
experiment. In the model of Ref. \cite{oh}, the $K^*$ 
exchange gives a large asymmetry, as large as 0.5 at 
$\cos\theta_{cm} \simeq$ 0.5-0.8 and $E_\gamma \leq 2.4$ GeV. 
Therefore, a small beam asymmetry suggests that $K^*$ 
exchange is somewhat suppressed. 
However, the discrepancy between the theoretical curve and the data in
Fig.~\ref{fig:asym} requires further theoretical studies.

In general, the experimental values are consistently more negative 
than the theoretical predictions 
over the entire photon energy range of 1.5-2.4 GeV, which are near 
to zero for the whole integrated angular range.
It is possible that final-state interactions in the deuteron 
target could change the theoretical calculations, and since the 
predicted beam asymmetry is small, these effects could be significant.
The present data will serve to constrain future improvements 
in the theoretical calculations.

In summary, we have measured the first photoproduction 
cross sections and beam asymmetries for $K^+\Sigma^{*-}$ 
photoproduction from the deuteron, by detecting both a 
$K^+$ and $\pi^-$ in the final state using the LEPS spectrometer. 
The measurement is limited to forward angles 
($\cos\theta_{cm} > 0.6$) and photon energies of 1.5-2.4 GeV. 
Simultaneously, \sigm~photoproduction was measured and shows 
good agreement with previous data \cite{kohri}.  The cross 
sections for \sigs~photoproduction peak at about 1.85 GeV 
and show a nearly-flat angular distribution. The
beam asymmetries are small and negative, and comparison with 
calculations in the model of Ref. \cite{oh} may suggest that there 
are multiple $s$-channel resonance contributions to \sigs 
photoproduction, and that $K^*$ exchange in the $t$-channel 
is largely absent.  
These data may also be useful to constrain other theoretical 
calculations of \sigs~photoproduction, such as in the model of 
Lutz and Soyer \cite{lutz}, and will be explored in a future 
paper.

The authors gratefully acknowledge the contributions of the staff at 
the SPring-8 facility.  This research was supported in part by the 
US National Science Foundation, the Ministry of Education, Science, 
Sports and Culture of Japan, the National Science Council of the 
Republic of China, and the KOSEF of the Republic of Korea.

\end{document}